\documentstyle[12pt]{article}

\begin{document}
\null\vskip -1cm
\centerline{
\vbox{
\hbox{September 1995}\vskip -9pt
\hbox{hep-ph/9509308}\vskip -9pt
     }
\hfill
\vbox{
\hbox{FERMILAB-PUB-95/300-T}\vskip -9pt
\hbox{CPP-95-14}\vskip -9pt
\hbox{DOE-ER-40757-071} \vskip -9pt
\hbox{UCD-95-29}\vskip -9pt
     }     } \vskip 1cm

\begin{center}
{\large \bf Color-Octet Quarkonium Production at the $Z$ Pole}

\vspace{0.1in}

Kingman Cheung\footnote{Internet address: {\tt cheung@utpapa.ph.utexas.edu}}

{\it Center for Particle Physics, University of Texas at Austin,
Austin, TX 78712}

\vspace{0.1in}

Wai-Yee Keung\footnote{Internet address: {\tt keung@fnalv.fnal.gov}}

{\it Physics Department, University of Illinois at Chicago, IL 60607-7059\\
Fermilab, P.O. Box 500, Batavia, IL 60510}
\vspace {0.1in}

Tzu Chiang Yuan\footnote{Internet address: {\tt yuantc@ucdhep.ucdavis.edu}}

{\it Davis Institute for High Energy Physics \\
University of California at Davis, Davis, CA 95616}

\end{center}

\begin{abstract}
The direct production rate of $J/\psi$
via the color-octet mechanism is calculated at the $Z$ resonance.
The color-octet production process $Z\to J/\psi q \bar q$
is shown to have a substantial branching ratio as well as  a
distinctive energy spectrum, which can be used as a powerful tool
to distinguish from the color-singlet direct production of the $J/\psi$.
\end{abstract}

\thispagestyle{empty}

Despite charmonium and bottomonium have been discovered some times ago,
heavy quarkonium physics has constantly drawn a lot of attentions from
theorists.  This is mainly due to the property of asymptotic freedom of QCD
which allows theorists to calculate perturbatively the short-distance part
in the production and in the decay rates of quarkonia, while the incalculable
long-distance part can be factored out and absorbed into the wave-function for
the ground state or its derivatives for higher orbital excitations.
Confinement implies that quarkonium must be seen as a color-singlet object.
The most natural assumption to make in the calculation of the
short-distance part of quarkonium production or decay rate
is that the heavy quark-antiquark pair inside the bound
state is produced or annihilated with the same orbital angular momentum,
spin, and color of the asymptotic physical quarkonium states.
This simple picture dubbed as the color-singlet
model \cite{schuler} in the literature is derivable from QCD and
has greater predictive power than other alternatives \cite{weiler}
like the color-evaporation model or the local duality approach.

During the past several years, a wealth of quarkonium data collected by the
CDF detector at the Tevatron has posted great challenge to the
naive color-singlet model.  With the advance of the technology of
silicon vertex detector, CDF in the 1992-1993 run was able to separate
their quarkonium data into two vast classes \cite{cdf}:
the first class can all be
traced back to a secondary displaced vertex which is a few microns
away from the interaction point and presumably they are all
coming indirectly from B meson decays; the second class of the
data was called prompt because it comes from direct QCD production.
Surprisingly,  the rates for the prompt $\psi$, $\psi'$, and $\chi_{cJ}$
at the large transverse momentum region were observed to be
orders of magnitude above the lowest order perturbative calculation
based on the color-singlet model \cite{cdf}.

To resolve these discrepancies one needs to seek for new production mechanisms
as well as new physical insights that go
beyond the color-singlet model.
The factorization model based on non-relativistic QCD (NRQCD)
recently advocated by Bodwin, Braaten, and Lepage \cite{bbl}
provides a new framework to calculate the inclusive production and
decay rates of quarkonium
to any order in strong coupling constant $\alpha_s$, as well as to any
order in $v^2$ where $v$ is the typical
relative velocity of the heavy quarks inside the bound state.
At the same time, the importance of fragmentation contributions
to the quarkonium production at the Tevatron was also pointed out by
Braaten and Yuan \cite{braaten-yuan}.
Parton fragmentation into quarkonium
is formally of higher order in strong coupling constant but
they can be enhanced over the traditional lowest order gluon-fusion
mechanism \cite{fusion} at sufficiently high transverse momentum.
However, cross sections for prompt $\psi$ and $\psi'$ production
were observed to be more than an order of magnitude above the predictions
based on the color-singlet model even after the inclusion of the fragmentation
contributions \cite{BDFM,CG,roy}.

One crucial feature of the factorization model is that quarkonium is not
solely regarded as simply a quark-antiquark pair but rather a superposition
of Fock states. In the light-cone gauge, the Fock state expansion
for the $\psi$ is
\begin{eqnarray}
\vert \psi \rangle &=& \vert c\bar c(^3S_1, {\underline 1} )\rangle +
{\cal O}(v) \;\vert c\bar c (^3P_J, {\underline 8})g \rangle
+ {\cal O}(v^2) \biggr(\; \vert c\bar c (^3S_1, \underline{8} \; {\rm or} \;
\underline{1} ) gg \rangle + \vert c\bar c(^1S_0,\underline{8})g \rangle
\biggr ) \nonumber \\
&+& {\cal O}(v^2) \; \vert c\bar c(^3D_{J'}, \underline{8} \; {\rm or} \;
\underline{1} )gg \rangle + {\cal O}(v^3) \; ,
\label{fock-psi}
\end{eqnarray}
where the angular momenta of the $c\bar c$ pair in each Fock state is labeled
by $^{2S+1}L_J$ with a color configuration of either $\underline{8}$ or
$\underline{1}$.
A novel idea proposed by Braaten and Fleming  \cite{braaten-fleming}
to explain the CDF data on prompt $\psi$ ($\psi'$) production  is due to
a color-octet term in the gluon fragmentation function into $\psi$ ($\psi'$).
This term corresponds to the case where the fragmenting gluon forms a
$c \bar c$ pair in the color-octet $^3S_1$ state at the short-distance
of order $1/m_c$ or smaller, and subsequently evolves non-perturbatively
into the physical $\psi$ or $\psi'$.
The non-perturbative effects are represented by the NRQCD matrix
element  $\langle O^{\psi}_8(^3S_1) \rangle$
($\langle O^{\psi'}_8(^3S_1) \rangle$) \cite{bbl}.
These color-octet matrix elements are suppressed by
order $v^4$ relative to the corresponding color-singlet
matrix elements $\langle O^{\psi}_1(^3S_1) \rangle$
($\langle O^{\psi'}_1(^3S_1) \rangle$). On the other hand, the
short-distance factor computed in perturbative theory  for this
color-octet process is only of order $\alpha_s(2m_c)$, which is enhanced
by a factor of $1/\alpha_s^2$ relative to the color-singlet process.
Therefore, the suppression in the color-octet matrix element can be
easily compensated by the enhancement in the  corresponding
short-distance factor when compared with the color-singlet process.
The CDF data for $\psi$ ($\psi'$) can be explained by including this
color-octet term in the gluon fragmentation function and by adjusting
the value for the matrix element  $\langle O^{\psi}_8(^3S_1) \rangle$
($\langle O^{\psi'}_8(^3S_1) \rangle$)  to fit the
data \cite{braaten-fleming,CGMP,cho-leibovich}.

While the color-octet mechanism can easily explain the transverse
momentum spectrum of the quarkonium measured at the Tevatron,
the normalization cannot be determined a priori.  One has to rely upon the
CDF $\psi$ and $\psi'$ data to extract the unknown
color-octet matrix elements.
In order to establish firmly our belief of the color-octet mechanism, it is
therefore important to identify these color-octet quarkonium signals in other
production processes. Recently,
Cho and Leibovich \cite{cho-leibovich}
have studied color-octet quarkonium production
at the Tevatron using exact tree-level calculations rather than the
fragmentation approximation.
This approach allows them to probe the CDF charmonium and bottomonium
data at almost the whole range of transverse momentum.
Braaten and Chen \cite{braaten-chen} showed that
the color-octet mechanism can give rise to a distinctive
signature at the upper endpoint of the energy
distribution of $\psi$ at CLEO energy.
In this Letter, we study the prompt quarkonium production by the color-octet
mechanism at the $Z$ resonance.
This work was motivated by the capability of the LEP detectors to separate
the prompt $\psi$ from those coming from B decays \cite{LEP}.
Besides, the preliminary results of prompt $\psi$ production at LEP \cite{LEP}
suggested almost a factor of ten larger than the prediction of the
most dominant color-singlet process via charm quark fragmentation
\cite{bck,cfrag}.
We shall show that the most important
color-octet process will give rise to a branching ratio
substantially larger than the most dominant color-singlet process
and is consistent with the preliminary LEP results \cite{LEP}.
Furthermore, the energy spectrum predicted by the color-octet production
process is very soft and should be distinguishable from that produced by
the most dominant color-singlet process.  Hence, our consideration of
$\psi$ production in $Z$ decays is crucial in verifying the
importance of the color-octet mechanism in quarkonium production.

According to the factorization formalism \cite{bbl}, the differential rate
for the inclusive production of a charmonium state $H$
with momentum $P$ from $Z$ decay can be written in a factored form:
\begin{equation}
\label{factorization}
d \Gamma (Z \to H(P) + X)  =
\sum_n d \hat \Gamma (Z \to c \bar c (P,n) + X)
\langle {\cal O}^H_n \rangle \; ,
\end{equation}
where $d\hat \Gamma$ is the partonic decay rate for producing a
$c\bar c$ pair with total momentum $P$, vanishing relative momentum, and
in an angular momentum and color state labeled collectively by $n$.
The long-distance factor $\langle {\cal O}^H_n \rangle$ is the
NRQCD matrix element describing the probability for the formation
of the quarkonium state $H$ from a $c \bar c$ pair in the state $n$.

The leading order color-singlet processes are
$Z \to \psi g g$ \cite{keung} and $Z \to \psi c \bar c$ \cite{bck}.
Although both processes are of order $\alpha_s^2$, the latter is two orders
of magnitude larger. The latter process has been interpreted as a
fragmentation process in which the $c \bar c$ pair was first produced
almost on-shell from the $Z$ decay and then followed by the fragmentation of
$c$ or $\bar c$ into the $\psi$ \cite{cfrag}.
Therefore, the process $Z\to \psi c \bar c$ is not suppressed by the quark
propagator effect of order $1/M_Z$ as it does in the process
$Z\to \psi gg$.
In the fragmentation approximation, the energy distribution of $\psi$ in
the process $Z\to \psi c\bar c$ is given by
\begin{equation}
\label{14}
\frac{d\Gamma}{dz}\left(Z\to \psi(z) \, c\bar c  \right) \approx
2\; \Gamma(Z\to c\bar c) \times D_{c\to \psi}(z) \;,
\end{equation}
where
\begin{equation}
D_{c\to \psi}(z) = \frac{16\alpha_s^2(2m_c) \langle {\cal O}_1^{\psi}(^3S_1)
\rangle }{243 m_c^3} \frac{z(1-z)^2}{(2-z)^6} \;
\left(16-32z+72z^2-32z^3+5z^4 \right ) \;,
\end{equation}
and $z=2E_\psi/M_Z$ with $E_\psi$ denotes the energy of $\psi$.
Both processes involve the same matrix element
$\langle {\cal O}_1^\psi (^3S_1) \rangle$ whose value can be extracted from
the leptonic width to be about $0.73 \; {\rm GeV}^3$.  Numerically,
the widths for $Z\to \psi gg$ and $Z \to \psi c \bar c$ are
about $6\times 10^{-7}$ GeV and $7\times 10^{-5}$ GeV, respectively.
There is also a higher order color-singlet process $Z\to q \bar q g^*$
followed by the gluon fragmentation $g^* \to \psi gg$ \cite{hagi}.
The branching ratio was estimated to be of order $10^{-6}$ only
\cite{braaten-yuan,hagi}.
However, the recent results from LEP \cite{LEP} showed a branching ratio
of order $10^{-4}$ for prompt $\psi$ production, which is
well above all the predictions from the color-singlet model.

The leading order color-octet process is of order $\alpha_s$ given by
the process $Z \to \psi g$, for which one of the Feynman diagrams is shown
in Fig.~1(a). It is understood that soft hadrons are around to match the
color.  The decay rate of this channel is given by
\begin{eqnarray}
\Gamma ( Z\to \psi g) &=& \frac{64\pi}{3} F g_v^2
\frac{\langle O_8^{\psi} (^1S_0) \rangle }{M_Z M_\psi} (1-\xi) +
 \frac{16\pi}{9} F g_a^2
\frac{\langle O_8^{\psi} (^3S_1) \rangle }{M_Z M_\psi}(1-\xi^2) \nonumber \\
&+& \frac{256\pi}{9} F g_v^2 \frac{\langle O_8^{\psi} (^3P_0) \rangle}
{M_Z M_\psi^3}  \biggr [
\frac{(1-3\xi)^2}{1-\xi} + \frac{6(1+\xi)}{1-\xi}
+ \frac{2(1+3\xi +6\xi^2)}{1-\xi} \biggr] \; ,
\end{eqnarray}
where $\xi=M_\psi^2/M_Z^2$, $F=\alpha\alpha_s/(x_w(1-x_w))$,
$g_v={1\over4}-{2\over3}x_w$, $g_a=-{1\over4}$,
and we have used a heavy quark spin symmetry relation
$\langle {\cal O}_8^\psi(^3P_J) \rangle \approx (2J+1)
\langle {\cal O}_8^\psi(^3P_0) \rangle$.  This
color-octet process has a very distinctive signature that the $\psi$ has an
energy equal one-half of the $Z$ mass and is recoiled by a hard gluon jet.
Unfortunately, the short-distance factors are suppressed by at least one
power of $1/M_Z$. Numerically, the width for $Z\to \psi g$ is of
order $10^{-7}$ GeV which renders this process useless at the $Z$ resonance.

The dominant color-octet process actually begins at the order $\alpha_s^2$
in the process $Z \to \psi q \bar q$, for which one of the Feynman diagrams
is shown in Fig.~1(b). Here $q$ represents $u,d,s,c$, or $b$.
The energy distribution of $\psi$ for the process $Z \to \psi q \bar q$
is calculated, in the limit $m_q = 0$, to be
\begin{eqnarray}
\label{exact}
\frac{d \Gamma}{d z}(Z  \to \psi (z) q \bar q)
&=&  \frac{\alpha_s^2(2m_c)}{18} \Gamma (Z \to q \bar q)
\frac{\langle O_8^{\psi} (^3S_1) \rangle }{m_c^3} \nonumber \\
&\times & \left[ \left( \frac{(z-1)^2 + 1}{z} + 2 \xi \frac{2 - z}{z} +
\xi^2 \frac{2}{z} \right) \log \left( \frac{z + z_L}{z-z_L} \right)
- 2 z_L \right] \; , \nonumber \\
& \; &
\end{eqnarray}
where $z=2E_\psi/M_Z$, $\xi=M_\psi^2/M_Z^2$, and $z_L = (z^2 - 4 \xi)^{1/2}$.
The physical range of $z$ is $2\sqrt{\xi} <z< 1+\xi$.
Similar expression was derived before for the process
$Z\rightarrow \psi l^+l^-$ in QED \cite{fleming,robin}.
The above exact result can be simplified
into the following form in the limit $\xi \to 0$:
\begin{equation}
\label{approx}
\frac{d \Gamma}{d z}(Z  \to \psi (z) q \bar q)
\approx \frac{\alpha_s^2(2m_c)}{18} \Gamma (Z \to q \bar q)
\frac{\langle O_8^{\psi} (^3S_1) \rangle }{m_c^3}
\left[ \frac{(z-1)^2 + 1}{z} \log \left( \frac{z^2}{\xi} \right)
- 2z \right]  \; .
\end{equation}
Following the same procedures in Ref.~\cite{fleming}, one can show that
the above limit corresponds to the sum of the fragmentation processes of
$Z \to q \bar q$ followed by $q \to \psi$ or $\bar q \to \psi$ and
$Z \to q \bar q g$ followed by $g \to \psi$.

Using the value for the color-octet matrix element
$\langle {\cal O}_8^\psi(^3S_1) \rangle \simeq 0.015 \; {\rm GeV}^3$
extracted from the fit to prompt $\psi$ data at the Tevatron
\cite{CGMP,cho-leibovich}, $m_c=1.5$ GeV, $M_\psi \approx 2m_c$, and
$\alpha_s(2m_c)=0.253$,
we show the energy distribution of $\psi$ for this process
in Fig.~2 (solid curve).
The decay width of $Z \to \psi q \bar q$
can also be obtained by integrating Eq.~(\ref{exact}) over
the physical range of $z$, and is given by
\begin{eqnarray}
{\Gamma (Z\to \psi q\bar q)\over \Gamma(Z\to q\bar q)}&=&
\frac{\alpha_s^2(2m_c)}{36} \frac{\langle O_8^{\psi} (^3S_1) \rangle }{m_c^3}
\bigg\{5(1-\xi^2)-2\xi\log\xi+\bigg[
 2\,\hbox{Li}_2\left({\xi\over 1+\xi} \right)
-2\,\hbox{Li}_2\left({ 1 \over 1+\xi}\right)
\nonumber\\
&\ &\quad\quad -2\log(1+\xi)\log\xi + 3\log\xi+\log^2\xi\bigg](1+\xi)^2\bigg\}
\\
&\simeq&    2.2\times 10^{-4}  \;.
\end{eqnarray}
Here Li$_2(x)=-\int_0^x{dt\over t}\log(1-t)$ is the Spence function.
Summing over all the quark flavors $(q=u,d,s,c,b)$, we obtain
the decay width
$\sum_q \Gamma(Z\to \psi q\bar q) \simeq 3.8\times 10^{-4}$ GeV
and the branching ratio
$\sum_q {\rm Br}(Z\to \psi q\bar q) \simeq 1.5 \times 10^{-4}$.
Assuming the dominant prompt $\psi$ production process to be
$Z\to q\bar q g^*$ with $g^* \to \psi+X$, in which according to color-singlet
model the off-shell gluon fragments into a $\psi$ plus two perturbative
gluons, DELPHI obtained the limit
${\rm Br}(Z\to q \bar q g^*; \; g^* \to \psi +X)
< 4.1 \times 10^{-4}$ \cite{LEP}.
Since the event topology of our color-octet process is similar to this one,
this limit should also be valid for our color-octet process.  Therefore,
our result is consistent with this data.

In Fig.~2 we compare the energy distributions of $\psi$ coming from the most
important color-octet process $Z\to \psi q\bar q$ using Eq.~(\ref{exact})
and the most dominant color-singlet process $Z \to \psi c \bar c$
using  Eq.~(\ref{14}).
For the color-octet process we have summed over all quark flavors,
$q=u,d,s,c,b$, in the final state.
The comparison in Fig.~2 shows a very spectacular difference between the
color-octet and color-singlet processes.
The spectrum for the color-octet process is very soft with a pronounced peak
at the lower $z$ end, while the spectrum for the color-singlet process is
rather hard because of the nature of heavy quark fragmentation.
Though the spectrum for the color-singlet process will be
softened somewhat by including higher order QCD corrections \cite{cfrag},
the spectacular difference remains unchanged.
Another striking result is that the color-octet spectrum dominates
over the color-singlet one for all values of $z$.
This indicates that this color-octet contribution should not be neglected!

The energy distributions for the $\psi'$ and $\chi_c$
coming from the color-octet and color-singlet processes are similar
to those illustrated above for the $\psi$. The importance of the
color-octet contributions can also be shown for these two cases as well.
The rates for direct $\psi'$ production are roughly scaled down
by a factor of 5 from those of the $\psi$ shown in Fig.~2.
Furthermore, the production of $\Upsilon$ by the color-singlet process
$Z\to \Upsilon b\bar b$ and the color-octet process $Z\to \Upsilon q\bar q$
are similar to that of $\psi$, but with smaller branching ratios.
Details will be presented elsewhere.

In closing, we have considered the color-octet quarkonium production
at the $Z$ resonance.
The dominant direct production of $\psi$ is shown to be the color-octet
process $Z \to \psi q \bar q$,  which is a few times larger than the
most dominant color-singlet production process $Z\to \psi c\bar c$.
Furthermore, the energy spectrum of the $\psi$ for the color-octet
mechanism has been shown to be very different from that of the color-singlet
process.   Therefore, measurements of the decay width and the energy spectrum
for the prompt $\psi$ production at LEP can further
confirm the importance of the color-octet mechanism.

This work was supported in part by the United States Department of
Energy under Grant Numbers DE-FG02-84ER40173, DE-FG03-93ER40757,
and DE-FG03-91ER40674.

{\em Note added}: After this paper was submitted, similar results were
reported by Cho \cite{cho}.


\section*{Figure Captions}
\begin{enumerate}

\item \label{fig1}
One of the contributing Feynman diagrams for the processes (a) $Z \to \psi g$,
(b) $Z\to \psi q\bar q $, and (c) $Z \to \psi c \bar c$.

\item \label{fig2}
The energy spectrum $d\Gamma/dz$ of the $\psi$ from the color-octet
process $Z\to \psi q \bar q$ and the color-singlet process
$Z \to \psi c \bar c$.

\end{enumerate}

\newpage
\input epsf

\begin{figure}
\centering
\leavevmode
\epsfysize=550pt
\epsfbox{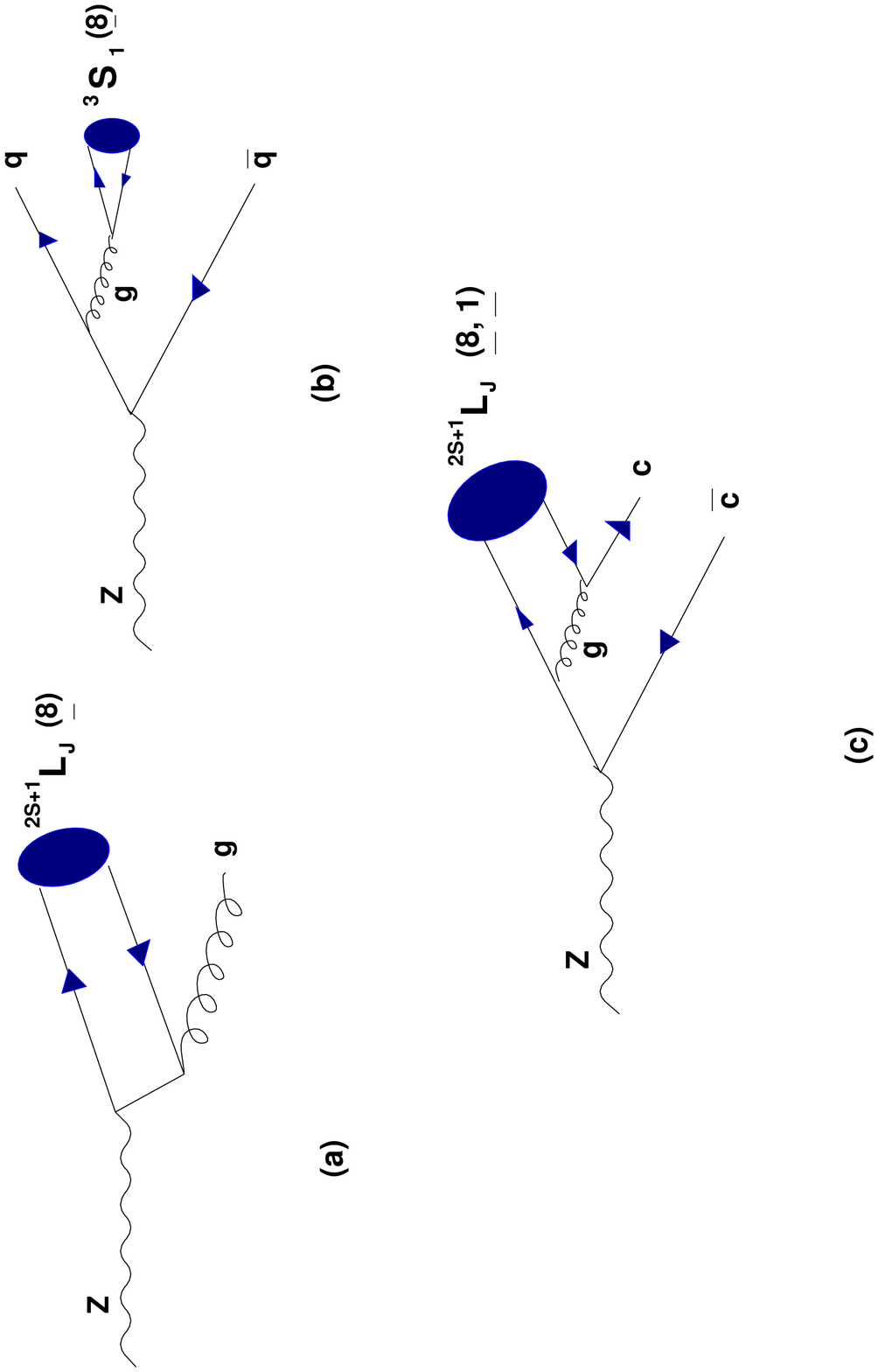}
\end{figure}

\begin{figure}
\centering
\leavevmode
\epsfysize=550pt
\epsfbox{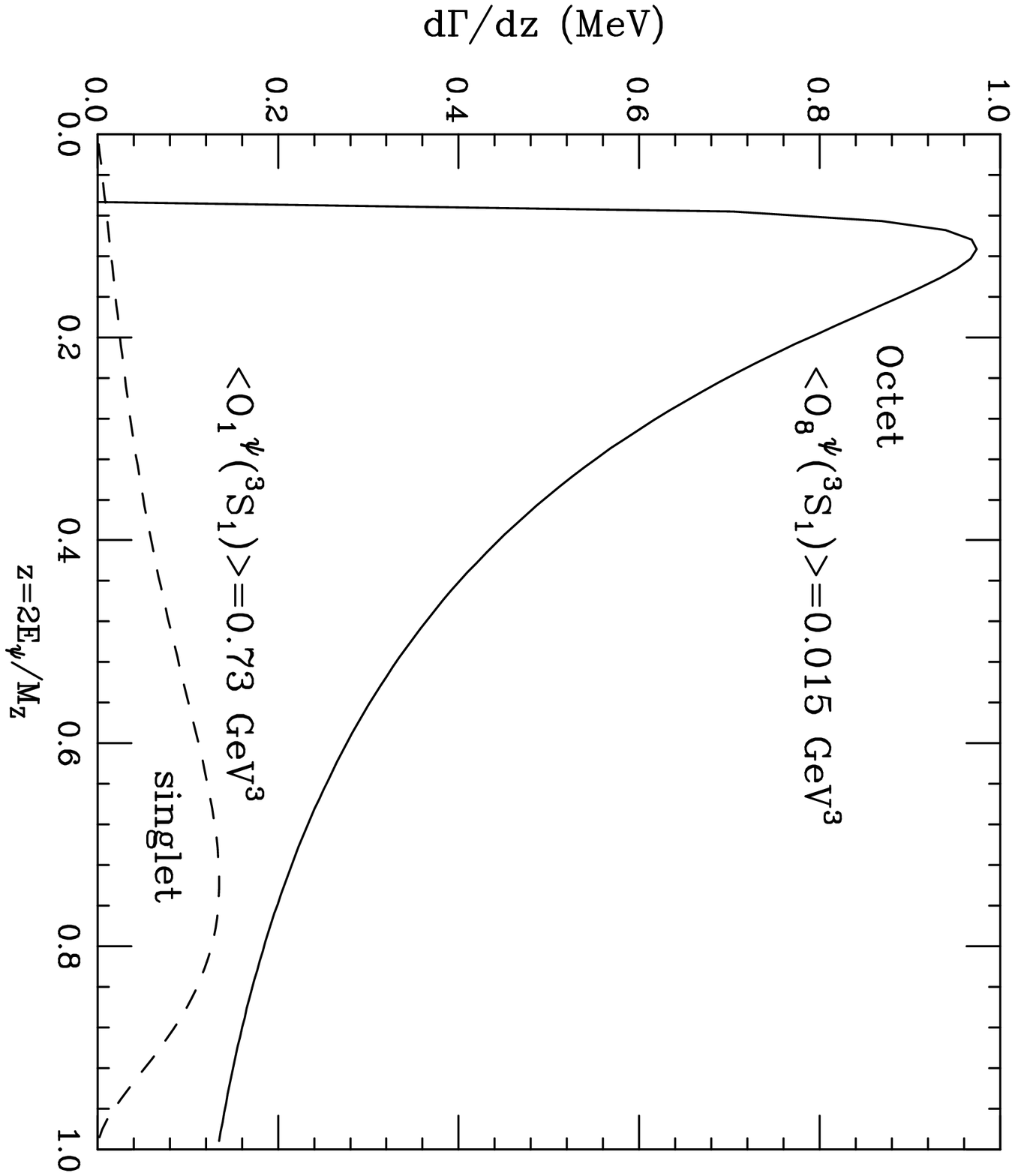}
\end{figure}

\end{document}